\documentclass[showpacs,preprint,amssymb]{revtex4}
\usepackage{graphicx}% Include Figure files
\usepackage{dcolumn}% Align table columns on decimal point
\usepackage{bm}% bold math
\def\be{\begin{equation}} 
\def\ee{\end{equation}}
\def\bea{\begin{eqnarray}} 
\def\eea{\end{eqnarray}}
\def\line{\hbox to \hsize}    
\def\frac #1#2{{#1\over #2}}

\def\det{{\rm det\,}}

\def\det{{\rm det\,}}

\def\1{\mbox{\bf 1}}
 
\def\levelonelist{
        \begin{list}{\mybulA}%
                        {
        \setlength{\topsep}{0pt}
        \setlength{\parsep}{0pt}
        \setlength{\partopsep}{0pt}
        \setlength{\itemsep}{0pt}
                        }
                }
\def\leveltwolist{
        \begin{list}{\mybulB}%
                        {
        \setlength{\topsep}{0pt}
        \setlength{\parsep}{0pt}
        \setlength{\partopsep}{0pt}
        \setlength{\itemsep}{0pt}
                        }
                }
\def\bo{\levelonelist}

\def\el{\end{list}}
\begin{document}

\title{Symmetries, Dimensions, and Topological Insulators:  the mechanism    behind the face of the Bott clock}

 \author{ MICHAEL STONE}

\affiliation{University of Illinois, Department of Physics\\ 1110 W. Green St.\\
Urbana, IL 61801 USA\\E-mail: m-stone5@illinois.edu}   

 \author{ CHING-KAI CHIU}

\affiliation{University of Illinois, Department of Physics\\ 1110 W. Green St.\\
Urbana, IL 61801 USA\\E-mail:chiu7@illinois.edu}

\author{ABHISHEK ROY}
\affiliation{
University of Illinois, Department of Physics\\ 1110 W. Green St.\\
Urbana, IL 61801 USA
\\E-mail Aror2@uiuc.edu}

\begin{abstract}  

We provide an account of some of the mathematics of Bott periodicity and  the Atiyah, Bott, Shapiro construction.   We apply these ideas to understanding the twisted bundles of  electron bands that underly  the properties of topological insulators, spin Hall systems, and other topologically interesting materials.

\end{abstract}


\maketitle

\section{Introduction}
\label{SEC:introduction}

Topological insulators  and superconductors are  many-fermion systems  possessing an unusual band structure that leads to a  bulk band  gap, but   topologically-protected  gapless  extended surface modes. The existence of such  materials was predicted theoretically \cite{kane1,bernevig1,moore,fu,roy1,roy2}, and  several  examples  have now been confirmed  experimentally \cite{hsieh1,hsieh2,xia,konig,chen_arpes}.  Part of the topological protection arises from generic symmetries of the of the underlying one-particle Hamiltonians. These   symmetries  include   time reversal and, in the case of superconductors,  the particle-hole symmetry  of   the Bogoliubov-de Gennes (BdG) Hamiltonian.   

There is a subtle interplay between the possibility of topologically non-trivial band structure, the symmetries, and the dimensions of the system \cite{hughes1,ludwig0,ludwig1,ludwig2}. This  interplay is  displayed in table \ref{TAB:lud}.  In this table the first two column contain the names associated with  the symmetry class in the  Dyson  scheme \cite{dyson}, as completed to include superconductivity by Altland and Zirnbauer \cite{zirnbauer1,zirnbauer2}.  The next three columns display the symmetries possessed by the Hamiltonians in this class. A minus sign indicates that the symmetry operation involves the electron spin  and so squares to minus the identity. The last four columns indicate whether a non-trivial topological phase is possible in $d$ dimensions. A ${\mathbb Z}_2$ indicates that there are two possibilities --- trivial or non trivial; a ${\mathbb Z}$ indicates that there are infinitely many  possible phases that are classified by an integer; a ``$0$'' means that no non-trivial topology can exist.  The data in this table  was obtained  in     \cite{hughes1,ludwig0}, but   it was Kitaev who first pointed out \cite{kitaev_K} the  striking pattern of correlations between the    symmetry, topology, and dimension. The correlations  only become manifest   after  the AIII and A classes are treated separately, and the remaining eight classes  are displayed  in a particular order. Kitaev  explained that the pattern  arises from the interplay between topological K-theory and  the  two-and  eight-fold  Bott  periodicity \cite{bott,milnor} of the homotopy groups of ${\rm U}(n)$ and ${\rm O}(n)$  respectively.

Topological  K-theory classifies  vector bundles.  At each point ${\bf k}$ of the Brillouin zone we have a Hamiltonian $H({\bf k})$,  and associated with it the subspace of  negative-energy eigenstates that is to be  filled by electrons.  If, as we explore the Brillouin zone, this subspace twists in a manner that cannot be continuously undone, then the bundle of vector spaces is {non-trivial}. The non-triviality will be reflected in physical effects such as a quantum Hall effect, chiral surface states, and the absence of localized Wanier functions.   A complete classification of which vector bundles can be deformed into one another is difficult. K-theory simplifies this classification by relaxing the notion of equivalence.  The resulting  lack of precision might seem like a bad thing,   but it is not. Our  desire  is only to classify bundles in such way as to capture their features that are  essential for the physics. For example, if we have a material  whose Fermi sea has  a single filled  band with Chern number $n$, it has the same Hall conductance as a solid possessing $n$ filled bands each with Chern number unity.   The corresponding vector bundles have different rank (dimension of the vector spaces)  and so cannot be deformed into each other.  In the reduced K-theory of bundles over a common base-space $X$, denoted  $\tilde K(X)$,  bundles of different rank  are  counted as    equivalent
if we can deform them into each other after  adding suitable trivial bundles.  In the  one-band versus $n$-bands example, once we add a rank  $n-1$ trivial bundle (for example $n-1$ previously ignored localized atomic core states)  to the the Chern-number $n$ band, it can be continuously deformed into the other bundle.  The bundles are equivalent  both in $\tilde K(X)$ and in their physical properties.

The full machinery of K-theory is intimidatingly abstract, but the simpler   mechanism   that underlies  the   period-two or period-eight  pattern of  correlations  can be understood with relatively unsophisticated  mathematical tools ---  representation theory and the basics of homotopy as described  in \cite{mermin}, or perhaps \cite{crossley}.  It  aim  of this present paper to explain how this mechanism   works, and to fill in some of details sketched in 
  \cite{kitaev_K}.

In the next section we will explain why the the symmetric spaces that form the basis of the Altland-Zirnbauer classification naturally occur in a certain order, and  relate these symmetric spaces to the Altland-Zirnbauer Hamiltonians. We  also  explain why the symmetric space that classifies the ground-state bundles is one step  further  round  the periodic Bott clock from the space that contains the evolution operators.   Section three  reverses the discussion of Altland and Zirnbauer and derives the symmetry classes from the symmetric spaces. Section four  is a slight digression that prepares some ingredients  that are needed to accommodate the  fact that time reversal reverses the direction of the Bloch momentum.  Section five  discusses the  Atiyah-Bott-Shapiro \cite{atiyah_clifford}  theory of real representations of Clifford algebras and in section six uses it  to construct Hamiltonians in any symmetry class and with any possible topologically twisted ground state.  

One thing that we do {\it not\/} do in this paper is discuss the complications that arise because the Brillouin zone is a torus. We restrict ourselves to constructing bundles over spheres. This is sufficient for the case of ``strong'' topological insulators, but it does not capture the possibility of ``weak'' (lower dimensional) insulators. 

\begin{table}
\begin{center}
\begin{tabular}{|c|c|c|c|c|c|c|c|c|}
\hline
Cartan & Dyson name  &TRS&PHS&SLS & $d=0$ & $d=1$& $d=2$ & $d=3$ \\
\hline\hline
AIII & chiral unitary& $0$&$0$&1& $0$ &${\mathbb Z}$ &$0$& ${\mathbb Z}$\\ 
A&  unitiary &$0$&$0$ &$0$ &${\mathbb Z}$ &$0$& ${\mathbb Z}$&$0$\\
\hline\hline
D&  BdG&${\phantom-} 0$&$+1$&${\phantom-}0$& ${\mathbb Z}_2$ & ${\mathbb Z}_2$& ${\mathbb Z} $& $0$\\ 
DIII&  BdG&$ -1$&$+1$& ${\phantom-}1$ & $0$ & ${\mathbb Z}_2$ &   ${\mathbb Z}_2$ & ${\mathbb Z}$ \\
AII & symplectic & $-1$& ${\phantom-}0$ &${\phantom-}0$&  ${\mathbb Z}$ & $0$& ${\mathbb Z}_2$&${\mathbb Z}_2$\\
CII & chiral symplectic & $-1$&$-1$& ${\phantom-}1$&  $0$ &${\mathbb Z}$ & $0$ &${\mathbb Z}_2$\\
C & BdG &$ {\phantom-}0$& $-1 $& ${\phantom-}0$&  $0$ &$0$ & ${\mathbb Z}$ &$0$\\
CI & BdG& $+1$&$-1$& ${\phantom-}1$& $0$ &$0$ & $0$&${\mathbb Z}$ \\
AI & orthogonal  &$ +1$&${\phantom-}0$&${\phantom-}0$& ${\mathbb Z}$ & $0$&$0$ & $0$  \\
BDI & chiral orthogonal & $+1$&$+1$&${\phantom-}1$& ${\mathbb Z}_2$&${\mathbb Z}$ &$0$& $0$\\
\hline\hline
\end{tabular}
\end{center}
\caption{The Dyson-Altland-Zirnbauer Hamiltonian classes, their symmetries, and  possible   topological phases.  (After table 4 in \cite{ludwig1}) }
\label{TAB:lud}
\end{table}

\section{The Bott  sequence of Symmetric Spaces}
\label{SEC:bott_sequence}

We begin with a somewhat backwards account of   the  Altland-Zirnbauer tenfold-way classification of quantum  Hamiltonians.  Altland and Zirnbauer proceed \cite{zirnbauer1,zirnbauer2} by considering families of single electron Hamiltonians with given  discrete symmetries --- time reversal or BdG particle hole symmetry --- each  with and without spin. They show that   for each set of symmetries there is a corresponding Lie algebra $\mathfrak g$, and a  decomposition of this algebra as ${\mathfrak g} ={\mathfrak m} + {\mathfrak h}$, where 
\be
[{\mathfrak h},{\mathfrak h}]\in {\mathfrak h},\quad [{\mathfrak h},{\mathfrak m}]\in {\mathfrak m},\quad [{\mathfrak m},{\mathfrak m}]\in {\mathfrak h}.
\ee
The matrices that generate the  time evolution ($i$ $\times$ the Hamiltonian) for this class    are then precisely the set  ${\mathfrak m}$.  Corresponding to the algebras ${\mathfrak g} $ and ${\mathfrak h}$ are  compact Lie groups $G$ and $H\subset G$. From them we obtain  a homogeneous space  $G/H$ that is naturally a Riemannian manifold. The commutation relations  of the algebras ensure that  for any point in $G/H$ there exists an  isometry  $\sigma$ that reverses  the directions of geodesics though the  point. These isometries  make $G/H$ into a symmetric space, and  {\'Elie Cartan} classified all possible symmetric spaces  in the 1920's  \cite{cartan}. The  Altland-Zirnbauer classification uses  of all   ten of the   compact symmetric spaces in which the matrices can have arbitrary large dimensions. In this sense  their  classification is exhaustive.

In contrast we  proceed  in the reverse direction. We  begin  with the symmetric spaces  and from them  extract  the Altland-Zirnbauer Hamiltonian families.  Only then do we uncover their symmetries. The advantage is that we  discover  that the symmetric spaces arise in a natural order. They are  the sequence of  order-parameter spaces  that arise as we progressively  break  a large symmetry by introducing more and more symmetry-breaking operators. As a bonus, the  matrices representing the symmetry-breaking operators will later   serve as  the building blocks of  model Hamiltonians that yield all possible  topologically non-trivial band structures.  After two or eight steps  the sequence of spaces repeats itself. This is Bott periodicy.

The periods of  two and eight are associated with   two distinct super-families  in the   Altland-Zirnbauer classification, a set of two associated with the unitary group and a  set  of eight associated with the real orthogonal group. We will focus on the latter as that  is the  most intricate.
The discussion is clearest if we  follow Dyson \cite{dyson} and work as much as possible with matrices with real entries. In this we again  depart from Altland and Zirnbauer who mostly use   complex matrices. 

The large symmetry we begin with is the group ${\rm O}(N)$. Here $N$ itself should be large. The mathematics literature usually takes a formal limit $N\to \infty$, but we  only require that  $N$  be large enough that low-dimensional exceptions can be disregarded.  For example,  the homotopy groups $\pi_n({\rm O}(N))$ become independent of $N$ once $n\le N+1$. (Similarly   $\pi_n({\rm U}(N))$ is independent of $N$ once $n\le 2N+1$.) 

To recover all eight symmetric spaces,  $N$ needs to be a multiple of $16$, so we consider  the action of   ${\rm O}(16r)$ on a vector space $V$ over ${\mathbb R}$ of dimension $16r$.  The  symmetry breaking operators  will be a set   $\{J_i: i=1,\ldots, k\}$   of mutually anti-commuting orthogonal complex structures acting on this space.  This language means that the    $J_i$ are orthogonal matrices that square to $-{\mathbb I}$  and obey  
\be
J_iJ_j+J_jJ_i=-2\delta_{ij}{\mathbb I}.
\ee 
They  therefore constitute a representation of the  Clifford algebra ${\rm Cl}_{k,0}$.  We  usually think of  representations of Clifford  algebras as  being  Dirac gamma matrices.  We will, however,  reserve the symbols $\gamma_i$ for {\it irreducible\/}  representations of the algebra.  The $J_i$ will usually  be highly reducible. 

Because they are    
orthogonal matrices, we have that  $J_i^T= J_i^{-1} =-J_i$, and  it is sometimes convenient to regard the skew-symmetric  $J_i$  as being elements of ${\mathfrak o}(16r)$, the Lie algebra of  ${\rm O}(16r)$.

 The subgroup of ${\rm O}(16r)$ that commutes with  $J_1$,
\be
O^T J_1 O=J_1,
\ee
is ${\rm O}(16r)\cap {\rm Sp}(16r, {\mathbb R}) \simeq{\rm U}(8r)$.   If we think of $J_1$ being a block diagonal matrix  
\be
J_1= {\rm diag}\left[  \left(\matrix{0&-1\cr 1& \phantom - 0}\right),\ldots, \left(\matrix{0&-1\cr 1& \phantom - 0}\right)\right],
\ee
then commuting with $J_1$ forces a $16r$-by-$16r$  orthogonal matrix to take the form of an   $8r$-by-$8r$ matrix  whose entries are two-by-two blocks in the form
\be
a \left(\matrix{1 &0\cr 0 & 1}\right)+b  \left(\matrix{0&  -1\cr  1& \phantom - 0}\right) \to a+ib
\label{EQ:aplusib}
\ee
As indicated, these entries  can be regarded as    complex numbers $a+ib$. The matrix $J_1$ itself  then acts as  $i$ $\times$ the $8r$-by-$8r$ identity matrix on an $8r$ dimensional complex vector space.  It is because $J_1$  assembles  a complex space out of a real space  that it  is  called a {\it complex structure\/}.

In a similar manner, the  subgroup of  ${\rm U}(8r)$ that commutes with  both $J_1$ and $J_2$ is  the unitary symplectic group $  {\rm U}(8r)\cap  {\rm Sp}(8r;\mathbb C) \equiv {\rm Sp}(4r)\simeq {\rm U}(4r, {\mathbb H})$.  To obtain the   quaternionic  unitary group $  {\rm U}(4r, {\mathbb H})$  we   gather   the $16r$  real vector components into sets of four,   and from each quartet   construct a quaternion   ${\bf x}= x_0+ x_1 {\bf i}+ {\bf j}(x_2+{\bf i} x_3)$. The unit quaternions ${\bf i}$, ${\bf j}$, ${\bf k}$ then act on  ${\bf x}$ from the { left\/} as multiplication into  the column vector $(x_0,x_1,x_2,x_3)^T$ of  the four-by-four matrices  
\be
{\bf i} =  \left(\matrix{i & \phantom - 0\cr  0& -i}\right), \quad 
{\bf j}= 
\left(\matrix{0&-1\cr 1& \phantom -0}\right) , \quad 
{\bf k} = \left(\matrix{0 &-i \cr- i &0}\right).
\ee
Here  $1$ and $i$ are  shorthand for the 2-by-2 sub-blocks
\be
1= \left(\matrix{1 &0\cr 0 & 1}\right), \quad  i=\left(\matrix{0&-1\cr 1& \phantom - 0}\right).
\ee
All three four-by-four real matrices are skew symmetric.
The  quaternionic-conjugate transpose of a  matrix with quaternion entries   therefore  coincides with the ordinary transpose of the  corresponding four-times-larger real matrix.
The mutually anticommuting complex structures that commute with the ${\bf i}$, ${\bf j}$ and  ${\bf k}$  matrices  are in this basis 
\be
J_1={\rm diag} \left[\left(\matrix{ i &0 \cr 0 & i}\right),\ldots,\left(\matrix{ i &0 \cr 0 & i}\right)\right],\quad J_2= {\rm  diag}\left[\left(\matrix{0&\phantom -\sigma_3\cr -\sigma_3&0}\right),\ldots,\left(\matrix{0&\phantom -\sigma_3\cr -\sigma_3&0}\right)\right].
\ee
These matrices represent   the  action on the column vector $(x_0,x_1,x_2,x_3)^T$ of multiplying the quaternions ${\bf i}$ and  ${\bf j}$ into  ${\bf x}$  from the {\it right\/}, and so naturally commute with the matrices representing ${\bf i}$, ${\bf j}$ and  ${\bf k}$, as they  describe  the effect  of quaternion multiplication from the {\it left\/}. 

As $J_2$ anticommutes with $J_1$, in the complex space in which $J_1\mapsto i\,{\mathbb I}$,  it behaves as an antilinear map  that squares to $-1$.  Consequently, in a  vector space over $\mathbb C$, an   antilinear  map
\bea
\chi(\lambda {\bf v})=\lambda^*\chi({\bf v})
\eea
 that obeys $\chi^2=-1$ is   called a {\it quaternionic structure\/}.

Now consider the subgroups that survive the introduction of yet more  $J_i$'s.  We closely follow \cite{milnor}, and     readers  who are prepared to take the results on trust  may skip over some tedious enumeration  to the resulting pattern of symmetry breaking displayed in Eqn.\ (\ref{EQ:pattern}).

After introducing  $J_3$ we   consider the operator $K=J_1J_2J_3$.  This  operator commutes with both $J_1$ and $J_2$, and  obeys  $K^2={\mathbb I}$.  Therefore $K$ possesses   two quaternionic eigenspaces $V_\pm$ in which  it takes the values $\pm 1$ respectively. Let the dimensions of these quaternionic spaces be $n_1$ and $n_2$, so that $n_1+n_2=4r$. The subgroup  of  
${\rm Sp}(4r)$ commuting with $J_1$, $J_2$, $J_3$, and so preserving this structure is  ${\rm Sp}(n_1)\times {\rm Sp}(n_2)$.  If we stop at this point, these dimensions  can be any pair such that $n_1+n_2=4r$, but  order to be able to continue and define a $J_4$ we will see that we need to  take  $n_1=n_2 =2r$.  

Next   introduce $J_4$ and let $L= J_3J_4$ . We have that $LK=-KL$, $L^2=-{\mathbb I}$,   and $L$ commutes with $J_1$ and $J_2$. $L$ therefore  preserves the quaternionic structure, and is a quaternionic isometry  from  $V_+$ to $V_-$. It is therefore an element of ${\rm U}(2r, {\mathbb H})\simeq  {\rm Sp}(2r)$ (and could not exist unless $n_1=n_2$). Conversely,  such an isometry can be used to define $L$ and hence  $J_4= J_3^{-1} L$.   The group preserving this structure is the diagonal subgroup  ${\rm Sp}(2r)$ of  ${\rm Sp}(2r)\times  {\rm Sp}(2r)$.

Now introduce $J_5$ and construct $M=J_1J_4J_5$ which has $M^2={\mathbb I}$ and commutes with $K$  and $J_1$. $M$ therefore acts within the  $V_+$ (or $V_-$) eigenspace of $K$ and divides it into two mutually orthogonal  eigenspaces $W_\pm$ with $W_-= J_2 W_+$   Conversely such a decomposition uniquely determines $J_5$.  Since $J_2$  interchanges the first and second blocks in  ${\bf x}= (x_0 + x_1{\bf i}) + {\bf j}(x_2 +x_3 {\bf i})$, the   quaternionic isomorphisms that preserve this decomposition must mix only $x_0$ with $x_1$ and $x_2$ with $x_3$  and so the quaternionic matrix entries  can contain only  $1$ and ${\bf i}$. This  subgroup can therefore  identified with  ${\rm U}(2r)$.

Now  introduce $J_6$ and set $N= J_2J_4J_6$ which commutes with $K$ and $M$, and therefore acts within either of $W_\pm$  and splits it into two mutually orthogonal eigenspaces $X_\pm$ such that $X_-=J_1X_+$.  Since $J_1$ acts as the two-by-two $i$,  the  subgroup preserving this structure cannot   have matrix entries with the quaternionic four-by-four matrix  ${\bf i}$, and it can be identified with ${\rm O}(2r)\subset {\rm U}(2r)$.

Now introduce $J_7$ and set $P=J_1J_6J_7$ which commutes with  $K$, $M$ and $N$ and splits $X_+$ into $\pm 1$ subspaces $Y_\pm$. These may have differing dimensions, but in order to have the possibility of introducing $J_8$ we must take the dimensions to be equal.  The group  preserving  this decomposition is therefore ${\rm O}(r)\times {\rm O}(r)$.

Finally we introduce $J_8$.  Now  the orthogonal transformation $Q=J_7J_8$ commutes with $K$, $M$ and $N$ but anticommutes with $P$. It is therefore an isometry mapping  $Y_+ \leftrightarrow Y_-$.  In order to preserve this structure we must take the diagonal subgroup ${\rm O}(r)$ of ${\rm O}(r)\times {\rm O}(r)$. 

The   progressive symmetry breaking has   lead to the nested sequence of  groups
\be
\ldots {\rm O}(16r)\supset {\rm U}(8r)\supset {\rm Sp}(4r)\supset  {\rm Sp}(2r)\times {\rm Sp}(2r)\supset {\rm Sp}(2r) \supset {\rm U}(2r)
\supset {\rm O}(2r) \supset {\rm O}(r)\times {\rm O}(r)\supset {\rm O}(r)\ldots
\label{EQ:pattern}
\ee
The sequence  can be extended to the left, and to the right when $r$ is a suitably large  power of two. The pattern repeats with period eight. In each cycle  $r$ increases or decreases  by a factor of 16. 
%\be
%\ldots {\rm O}(16r)\supset {\rm U}(8r)\supset {\rm Sp}(4r)\supset  {\rm Sp}(2r)\times {\rm Sp}(2r)\qquad  \be
%\be
%\qquad \qquad \qquad\quad  \supset {\rm Sp}(2r) \supset {\rm U}(2r)
%\supset {\rm O}(2r) \supset {\rm O}(r)\times {\rm O}(r)\supset {\rm O}(r)\ldots
%\be

When complex numbers are allowed, the   corresponding sequence of groups is simpler. We can now multiply the $J_i$ by $i$ so that 
$K_i=iJ_i$ obeys $K_i^2={\mathbb I}$ and splits ${\mathbb C}^{2r}$ into two spaces that may have arbitrary dimensions. To keep going, however, we must take the dimensions  to be equal. Thus, we get
\be
\ldots {\rm U}(2r)\supset {\rm U}(r)\times {\rm U}(r) \supset {\rm U}(r)\ldots
\ee

We now motivate the construction of the symmetric spaces from the sequences of groups.
Suppose we  already possess   a  set  $\{J_1,\ldots, J_d\}$ and wish to add a    $J_{d+1}$ that anticommutes with them. Then the set of choices  for $J_{d+1}$ is parametrized by a symmetric space. To see this, we     first argue    that  the choices  are   parameterized by a homogeneous space and then prove   that the homogeneous space   obeys the stronger condition of being  symmetric.   
Let the subgroup of $G_0\equiv {\rm O}(16r)$ that commutes with $J_1, \ldots, J_i$,  be $G_i$ and its Lie algebra ${\mathfrak g}_i$. Then if $J_{i+1}$ squares to $-{\mathbb I}$ and anticommutes with  $J_1, \ldots, J_i$, so does $g^{-1}J_{i+1}g$ for any $g\in G_i$. The subgroup of $G_i$ that continues to commute with the new $J_{i+1}$   is $G_{i+1}$, and so it is reasonable to conjecture that the range of choices for $J_{i+1}$ is the orbit of $J_{i+1}$ under the action of conjugation by $G_{i} $, --- {\it i.e.\/}\ the  coset,  or homogeneous space $G_{i}/G_{i+1}$.  These cosets, together with  the two families of cosets  that arise  in the complex case, are displayed in the tables \ref{TAB:real_cartans} and 
\ref{TAB:complex_cartans}.  

 \begin{table}
\begin{center}
\begin{tabular}{|c|c|c|c|c|}
\hline\hline
Cartan label & Name & $R_q$ &$G/H$ & $\pi_0(R_q)$\\
\hline
D & BdG & $R_1$ &$ {\rm O}(16r)\times {\rm O}(16r)/{\rm O}(16r) \simeq {\rm O}(16r)$& ${\mathbb Z}_2$\\
DIII & BdG  & $R_2$& ${\rm O}(16r)/ {\rm U}(8r)$ & ${\mathbb Z}_2$\\
AII  & symplectic & $R_3$ &$  {\rm U}(8r)/{\rm Sp}(4r)   $ & $ 0$  \\
CII  &  chiral symplectic & $R_4$ & $ \{{\rm Sp}(4r)/{\rm Sp}(2r)\times {\rm Sp}(2r)\}\times {\mathbb Z}   $ &${\mathbb Z}$ \\
C & BdG &  $R_5$ &$ {\rm Sp}(2r)\times {\rm Sp}(2r) / {\rm Sp}(2r)\simeq {\rm Sp}(2r)   $&   $ 0  $  \\
CI  & BdG & $R_6$ &$ {\rm Sp}(2r)/{\rm U}(2r)  $& $ 0 $ \\
AI  & orthogonal & $R_7$ &$  {\rm U}(2r)/{\rm O}(2r) $& $  0 $ \\
BDI & chiral orthogonal & $R_0$ &$\{{\rm O}(2r)/{\rm O}(r)\times {\rm O}(r) \}\times {\mathbb Z}  $& $ {\mathbb Z} $\\
% D& $R_1$ & $ {\rm O}(r)\times {\rm O}(r)/{\rm O}(r)\simeq {\rm O}(r) $ & ${\mathbb Z}_2 $\\
\hline\hline
\end{tabular}
\end{center}
\caption{The sequence of coset spaces parametrizing the choices for successive $J_i$ possessing real entries.  The first column contains label of the space  in the Cartan classification. The second  is  the  Altland-Zirnbauer class whose time-evolution opertors realize $G/H$.  The large $N$ versions of these spaces serve as classifying spaces denoted by $R_q$   In this r{\^o}le the real Grassmannian is denoted by $R_0$.  The $\pi_0(R_q)$ column  displays  the set parametrizing the  disconnected pieces of $R_q$. }
\label{TAB:real_cartans}
\end{table}
\medskip 
\begin{table}
\begin{center}
\begin{tabular}{|c|c |c|c|c|}
\hline\hline
Cartan label & Name& $C_q$&  $G/H$ & $\pi_0(C_q)$\\
\hline
A & unitary &$C_1$&  $ {\rm U}(2r)\times {\rm U}(2r)/{\rm U}(2r) \simeq {\rm U}(2r)$&$ 0$ \\
AIII&chiral unitary&  $C_0$ & $\{{\rm U}(2r)/ {\rm U}(r)\times {\rm U}(r)\}\times {\mathbb Z}$ & ${\mathbb Z}$ \\
\hline\hline
\end{tabular}
\end{center}
\caption{The corresponding sequence of coset spaces parametrizing the choices for successive $J_i$ when complex numbers are allowed.}
\label{TAB:complex_cartans}
\end{table}
\medskip 

 We  say ``conjecture'' because it is not immediately clear that the action of $G_i$ is transitive.  It is easy to see that the orbit captures the connected part of the space of choices. The disconnected parts --- which are the parts of interest for Bott periodicity  --- need a bit more work, but the claim is correct.  For example, the  ${\mathbb Z}_2$  appearing as  $\pi_0(R_2)$ arises from the two disconnected  parts of the choice-space  for $J_1$ in which the   pfaffian ${\rm Pf}(J_1)$ takes the value $\pm 1$. Since 
 \be
 {\rm Pf}(M^TAM) = {\rm Pf}(A)\,\det (M),
 \ee
 each disconnected part is  accessible from the  another  {\it via\/}  conjugation by  an orthogonal matrix  $g$ with  $\det g=-1$. 
 The $\times {\mathbb Z}$ factors in the tables represent the choices of dimensions  in ${\rm O}(n_1)\times {\rm O}(n_2)$ and ${\rm Sp}(n_1)\times {\rm Sp}(n_2)$.
They make $\pi_0$ of these spaces equal to ${\mathbb Z}$.  

Now we show that spaces $G_{i}/G_{i+1}$ are in fact symmetric spaces.
If $a\in {\mathfrak g_i}$ then so is $J_{i+1}a J_{i+1}^{-1}$. Thus ${\rm Ad}(J_{i+1}) :{\mathfrak g_i}\to {\mathfrak g_i}$ is an  involutive (squares to the identity) automorphism of ${\mathfrak g_i}$. Let the eigenspaces of this map with eigenvalues $\pm 1$ be ${\mathfrak h}_i$ and ${\mathfrak m}_i$. Then ${\mathfrak g}_i= {\mathfrak h}_i\oplus  {\mathfrak m}_i$ as a vector space, and the automorphism property requires that
\be
[{\mathfrak h}_i,{\mathfrak h}_i]\in {\mathfrak h}_i,\quad [{\mathfrak h}_i,{\mathfrak m}_i]\in {\mathfrak m}_i,\quad [{\mathfrak m}_i,{\mathfrak m}_i]\in {\mathfrak h}_i.
\ee
Since ${\mathfrak h}_i$ is precisely  $ {\mathfrak g}_{i+1}$, this confirms  that the homogeneous space $G_i/G_{i+1}$ is indeed  a symmetric space.   Its connected part is ${\rm Exp}({\mathfrak m}_i)$.    Recall that  matrices   $m\in {\mathfrak m}_i$  are the  evolution generators   ($i$ $\times $ the  Hamiltonian) in the Altland-Zirnbauer families. They  commute with $J_1,\ldots, J_i$, and anticommute with $J_{i+1}$. 

We next  turn to  the topology of the symmetric spaces. The large-enough $N$ versions   of the $G_i/G_{i+1}$  become the classification spaces  known as  $R_{i+2}$, and  these spaces have homotopy groups that are independent of $N$.  The    bundle  classification we seek requires us to know all the  homotopy groups of the $R_q$, not just the $\pi_0(R_q)$ that are displayed  in table \ref{TAB:real_cartans}.
We therefore sketch the argument that leads to the key result  $\pi_n(R_m)\simeq \pi_{n+1}(R_{m-1})$. From this we can obtain all the  $\pi_n(R_q)$, and as a corollary  deduce Bott periodicity: that $\pi_{n+8}({\rm O}(N))\simeq \pi_n({\rm O}(N))$ once $N$ is large enough.  
 
We first show that each  symmetric space $R_{i+2}\equiv G_i/G_{i+1}$ in the Bott sequence of spaces is naturally embedded (as a totally geodesic submanifold) in the preceding one. To see this   let $A_i= J_{i}^{-1} J_{i+1}$. Then $A_i$ is a skew symmetric orthogonal matrix that squres to $-{\mathbb I}$.  It anticommutes with $J_{i}$ but {\it commutes\/}  with $J_1,\ldots, J_{i-1}$.  Thus $A_i\in {\mathfrak m}_{i-1}$. Now 
\be
J_{i+1}= J_{i} A_i
\ee
and 
\be
\gamma(t)= J_i \exp \{ \pi A_i t\}= J_i\cos \pi t+ J_{i+1}\sin\pi t
\ee
is a  geodesic in   $G_{i-1}/G_{i}$ that interpolates between $J_i=\gamma(0)$ and $-J_i=\gamma(1)$, and has $J_{i+1} =\gamma(1/2)$.  The set of  these geodesics is therefore parametrized by $G_{i}/G_{i+1}$.  This is just as the set of geodesics from the noth pole to the south pole of the sphere is parametrized by points on the equator of the sphere.
 Milnor \cite{milnor} shows that   these geodesics capture the topology of the loop space  $\Omega [G_{i-1}/G_{i}]$ in that  the homotopy groups of $\Omega [G_{i-1}/G_{i}]$   coincide with those of $G_{i}/G_{i+1}$ once $N$ is large enough.  Then the standard  isomorphism 
$\pi_n(\Omega X)\simeq \pi_{n+1}(X)$ gives us  $\pi_n(R_m)\simeq \pi_{n+1}(R_{m-1})$.  Alternatively we can regard the 
space swept out by the geodesics as homeomorphic to   the reduced suspension $\Sigma [G_{i}/G_{i+1}] $ of $G_{i}/G_{i+1}$.  Bott then shows \cite{bott} that this suspension captures enough of  $G_{i-1}/G_{i}$ that we obtain the same isomorphism. 
The two approaches are related because of the natural identification \cite{crossley}
\be
{\rm Map}_*(\Sigma X, Y)={\rm Map}_*(X,\Omega Y)
\ee
between basepoint preserving maps from the reduced suspension of $X$ to $Y$ and from $X$ to  the space of based loops in  $Y$.

The reason we need need the homotopy groups is because, as we said earlier,  the  $R_q$, $q=i+2$, serve as   {\sl classification spaces\/}  in the sense of bundle theory.  In  particular $R_0\equiv BO\times {\mathbb Z}$  is the  classifying space for real vector bundles.
Any rank-$n$ real vector bundle can be obtained as the pull-back of the tautological  bundle  over some  real grassmanian ${\rm Gr}_{n}({\mathbb R}^{n+m})$  \hbox{$\simeq{\rm O}(n+m)/{\rm O}(n)\times {\rm O}(m)$} consisting of $n$-dimensional subspaces of ${\mathbb R}^{n+m}$.  Here {\it tautological\/}  means that the   fibre over the point $p\in  {\rm Gr}_{n}({\mathbb R}^{n+m})$ is the  corresponding $n$-dimensional subspace $p \subset {\mathbb R}^{n+m}$.   For $m$ sufficently large,  the homotopy equivalence classes of the vector bundles correspond one-to-one with  the homotopy classes  of the maps from $X$ to ${\rm Gr}_{n}({\mathbb R}^{n+m})$. The ``large enough $m$'' limit  of ${\rm Gr}_{n}({\mathbb R}^{n+m})$ is denoted by $BO(n)$ and rank-$n$ real  bundles are classified by the homotopy classes $[X, BO(n)]$ 
of continuous map from $X$ to  $BO(n)$. In $\tilde K$-theory we relax the notion of bundle equivalence and identify bundles that become equivalent   when trivial bundles (flat bands) of any rank are added to them. In this case we forget $n\in {\mathbb Z}$  and $BO(n)$ becomes $BO$.  We then have  $\widetilde{KO}(X) =  [X, BO]$ the homotopy classes of maps from $X$ to $BO$.  We cannot forget the ${\mathbb Z}$ factors when commuting homotopy, however.  In particular  $\pi_8(BO)={\mathbb Z}\times BO$.
Similarly for complex bundles $\pi_2 (BU)={\mathbb Z}\times BU$.

 In  the application to topological insulators, we want a tighter classification than that given by maps into $BO$.  We are interested in the bundle  of negative energy eigenstates of a family of  Hamiltonians  $H({\bf x}) $ in a {\it given Altland-Zirnbauer class\/}. So we allow only smooth deformations of the Hamiltonian that remain in that class.  For the topological effects it is only the eigenstates that matter, and  not their energy. So, following Kitaev, we flatten the spectrum and  seek   $Q_i \in {\mathfrak m}_i$  that has eigenvalues $\pm i$ (recall that Zirnbauer's generators are $i$ times the Hamiltonian).  The  $Q_i$ are then in one-to-one correpondence with the negative-energy spaces of the original Hamiltonians.   If our family of Hamiltonians in ${\mathfrak m}_i$ is  parametrized by ${\bf x}$ in a space  $X$ (a Brillouin zone say) then the bundle of ground states over $X$ will be trivial if and only if the   homotopy class of maps from $X$ into whatever  classifying space  parametrizes the possible $Q_i$ contains the constant map.  
 
 To find this classifying space, observe that the matrices  $A_{i+1}$ that we met earlier all lie in  ${\mathfrak m}_{i}$ and square to $-{\mathbb I}$.  They   therefore can be used as a  $Q_{i}$. The converse is also true. If $Q_{i}\in {\mathfrak m}_{i}$ and $Q_{i}^2=-{\mathbb I}$,  then $J_{i+1}Q_{i}$  anticommutes with $J_1,\ldots, J_{i+1}$   and so is a candidate  $J_{i+2}$. The set of possible $Q_{i}$  is therefore parametrized by the    the set of choices, $G_{i+1}/G_{i+2}$ for $J_{i+2}$. That is by $R_{i+3}$.  We conclude that the classifying space for the   class  of Hamiltonians  in ${\mathfrak m}_i$, whose  evolution operators lie in $R_{i+2}$   is   $R_{i+3}$ --- the  next space  along  in the Bott clock. The set of  distinct ground-state bundles over $X$ is therefore  given by the homotopy classes  
$
[X,R_{i+3}]
$
of continuous maps from $X$ to $R_{i+3}$. In particular, when  $X$ is the  sphere $S^d$  these classes are given by $\pi_d(R_{i+3})$. Because  $\pi_n(R_m)\simeq \pi_{n+1}(R_{m-1})$,  this homotopy group  has the same number of distinct elements as $\pi_0(R_{i+3+d})$.   

Unfortunately, in the case of most interest, when  $X$ is a Brillouin zone, the above count  is not correct.  It would lead to the  stripes of $\mathbb Z$'s and ${\mathbb Z}_2$'s  in table \ref{TAB:lud}  sloping  the wrong way.  The correct result is  $\pi_0(R_{i+3-d})$. The change of sign of $d$ arises  because antilinear  symmetries  such as time reversal have the effect of inverting  the Bloch wavevector ${\bf k}$ to $-{\bf  k}$ modulo reciprocal lattice vectors.  The resulting bundles are therefore more intricate than those classified by simple homotopy. Their classification  is the subject of KR theory \cite{atiyah_KR}, and we will delay describing what happens until section  \ref{SEC:abs_construction}.

In \cite{kitaev_K}  Kitaev mentions the spaces $\widetilde{KO}^{-q}(X)$.
Topologists  define $\tilde K^{-1}(X) =\tilde K(\Sigma X)$ by analogy with cohomology theory where $H^n(X)=H^{n+1} (\Sigma X)$.  In this language 
 \be
\widetilde{KO}^{-q}(X) =  [\Sigma^q X, BO]= [X, \Omega^q BO]= [X, R_q].
\ee
When $X$ is $S^d$, we have
\be
  \widetilde{KO}^{-q}(S^d)= [S^d, R_q]= \pi_0(R_{q+d})
\ee
so $q$ should be $i+3$ for the Altland-Zirnbauer Hamiltonians in  ${\mathfrak m}_i$.

 \section{The discrete symmetries}
 \label{SEC:symmetries}

The Altland-Zirnbauer classes are usually and most simply characterized by the  presence or absence of  three  discrete symmetries \cite{ludwig0}. These are  a sublattice symmetry   (SLS) generated by a  linear   map   $P$ that anticommutes with the Hamiltonian $H$: 
\be
PH = -HP,
\ee
and  two  antilinear maps. The first,  ${\mathcal C}$, implements  BdG particle-hole symmetry (PHS). The second,  ${\mathcal T}$,   implements  time-reversal symmetry (TRS).    The antilinear   maps give
\be
{\mathcal C}H{\mathcal C}^{-1}=- H,\quad {\mathcal C}^2= \pm {\mathbb I};\qquad 
{\mathcal T}H{\mathcal T} ^{-1}=H,\quad {\mathcal T}^2 =\pm {\mathbb I}.
\ee
In any given  basis we can represent $\mathcal C$ and $\mathcal T$ by complex matrices $C$ and $T$ such that
\be
CH^*C^{-1}= -H, \quad C^*C= \pm {\mathbb I},\qquad TH^*T^{-1} = H, \quad T^*T=\pm {\mathbb I}.
\ee
In a vector space $V$ over ${\mathbb C}$, however, the operation of  complex conjugation is not a basis-independent notion.  To describe it in a basis-independent manner we need to  introduce a {\it real structure\/}. This is a map $\varphi$ that is antilinear 
\be
\varphi (\lambda v) =\lambda^* \varphi (v)
\ee
and obeys $\varphi^2={\mathbb I}$.  We can then decompose $V= W \oplus_{\mathbb R} i W$ where 
\be
W=\{v\in V : \varphi (v)=v\}, \quad  iW= \{v\in V : \varphi (v)=-v\}.
\ee
In effect $\varphi$ selects  privileged basis vectors   ${\bf e}_n$ that are counted as real. They span $W$. The set $i{\bf e}_n$ spans $iW$. A complex vector  ${\bf v}= (u_n+iv_n){\bf e}_n$ is decomposed into a real vector $u_n{\bf e}_n +v_n(i{\bf e}_n)$ of twice the dimension.  To recover the original complex space from this twice-as-big real space,  we need 
a complex structure $J$   such that the antilinearity of $\varphi$  corresponds to $\varphi J=-J\varphi$.   

If an operator commutes with a  real structure $\varphi$, then there exists a basis (the $e_n$ from above) in which the matrix representing the operator becomes real. 
 
 Now we find the symmetries possessed corresponding to each  evolution symmetric  space:

\bo

\item[D$\equiv{\rm O}(16r)$:]  The coset generators  $m\in {\mathfrak m}_{-1}$ are real skew-symmetric matrices. These are not diagonalizable within the reals. We need to  double the  the Hilbert space to ${\mathbb R}^{32r}$  and tensor with ``$i$''$=-i\sigma_2$ so that the Hamiltonian becomes $H= -i\sigma_2\otimes m$. Then taking  $\varphi = \sigma_3\otimes {\mathbb I}$ to be the real structure that defines complex conjugation, we have $\varphi H=-H\varphi $. This $\varphi$ therefore defines a particle-hole symmetry that squares to $+{\mathbb I}$.  We should not count the eigenvectors $v$ and $ iv\equiv  (- i\sigma_2 \otimes {\mathbb I})v$ as being distinct.  (We can regard  the ${\mathbb R}^{32r}$ space and the operators ``$i$'' and $\varphi$ as being inherited from the previous cycle of the Bott clock.)

 \item[DIII $\equiv {\rm O}(16r)/{\rm U}(8r)$:] The $m\in {\mathfrak m}_0$'s are real skew symmetric matrices that anticommute with $J_1$. We can keep ${\mathcal C}=\varphi$ as a particle-hole symmetry and take ${\mathcal T}= \varphi \otimes J_1$ as a time reversal that commutes with $H= - i\sigma_2\otimes M$ and squares to $-{\mathbb I}$.  
The product of ${\mathcal C}$ and ${\mathcal T}$  is $J_1$, and this a linear  (commutes with  $- i\sigma_2 \otimes {\mathbb I}$)  ``$P$'' type symmetry  that anticommutes with $H$.

\item[AII$\equiv {\rm U}(8r)/{\rm Sp}(4r)$:]  The generators $m\in {\mathfrak m}_1 $ are real skew matrices that commute with $J_1$ and anticommute with $J_2$. They can be regarded as skew-quaternion-hermitian matrices with complex entries.  We no longer need  set $i\to  - i\sigma_2 \otimes {\mathbb I}$ as the matrices no longer have elements coupling between the artificial copies. We instead  use $J_1$ as the surrogate for ``$i$.'' Now  
$H=J_1 m$ is real symmetric, and  commutes with ${\mathcal T}=J_2$. This ${\mathcal T}$  acts as a time reversal operator squaring to $-{\mathbb I}$.
   
\item[CII$\equiv {\rm Sp}(4r)/{\rm Sp}(2r)\times {\rm Sp}(2r)$: ]  The matrices $m\in {\mathfrak m}_2$ commute with $J_1$ and $J_2$  but anticommute with $J_3$. Again $H=J_1 m$.   We can set  ${\mathcal T}=J_3$ as this commutes with with $H$ and squares to $-{\mathbb I}$.  $P= J_2J_3$ anticommutes with $H$ but commutes with $J_1$ (and so is a {\it linear} map) while ${\mathcal C}= J_2$  anticommutes with $H$, is {\it antilinear\/} and squares to $-{\mathbb I}$.  

\item[C$\equiv \{{\rm Sp}(2r)\times {\rm Sp}(2r)\}/{\rm Sp}(2r)\simeq {\rm Sp}(2r)$:] The matrices $m\in {\mathfrak m}_3$ commute with $J_1$, $J_2$, $J_3$, and anticommute with $J_4$, and we can  restrict ourselves to the subspace in which  $K=J_1J_2J_3$ takes a definite  value, say $+1$.  
The Hamiltonian $J_1 m$ commutes with $J_4$ -- but $J_4$ does not commute with $J_1J_2J_3$ and so is not allowed as an operator on our subspace.  Indeed no product involving $J_4$ is allowed.  But  ${\mathcal C}=J_2 $ commutes with   $J_1J_2J_3$  and still anticommutes with $H$. Thus we still have a particle-hole symmetry squaring to $-1$.  The old time reversal $J_3$ now {\it anticommutes\/}  with $H$  and looks like another particle-hole symmetry, but is not really an independent one as in this subspace $J_3 = J_2J_1$ and $J_1$ is simply multiplication by ``$i$.''  

\item[CI$\equiv {\rm Sp}(2r)/{\rm U}(2r)$: ]  The $m\in {\mathfrak m}_4$ anticommute with $J_5$.    Now $J_4J_5$ commutes with $J_1J_2J_3$, and so is an allowed operator.  It   anticommutes with $H= J_1 m$ and commutes with $J_1$. It is therefore a ``${P}$''-type linear map.    The map ${\mathcal T}= J_2J_4J_5$ is antilinear (anticommutes with $J_1$), commutes with $H$ and   ${\mathcal T}^2=+{\mathbb I}$. We can take ${\mathcal C}=J_2$  again.  
 
\item[AI $\equiv {\rm U}(2r)/{\rm O}(2r)$: ]  The $m \in {\mathfrak m}_5$ n anticommute with $J_6$, and we are to restrict ourselves to the subspace on which $K=J_1J_2J_3=+1$ and $M=J_1J_4J_5=+1$.    The map ${\mathcal T}=J_3J_4J_6$    commutes with $K$ and $M$
and commutes $H=J_1 m$.    We have ${\mathcal T}^2=+{\mathbb I}$.   We could equivalently  take  ${\mathcal T}=J_2J_4J_6$.
 
\item[BDI$\equiv {\rm O}(2r)/{\rm O}(r)\times{\rm O}(r)$: ] The $m \in {\mathfrak m}_6$  anticommute with $J_7$, and we are to restrict ourselves to the eigenspaces of  $K=J_1J_2J_3$, $M=J_1J_4J_5$.  The $m$ also commute with the antilinear operator  $N=J_2J_4J_6$ which we can regard as our real structure $\varphi$.  We therefore set ${\mathcal C}=\varphi=N$.  Now $J_3J_4J_7$,  $ J_3 J_5 J_6$, $J_2J_5J_7$  and $J_1J_6J_7$   all commute with $K$, $M$ and $N$, each squaring to $+{\mathbb I}$. In the restricted subspace $J_3J_4J_7\propto J_2J_5J_7 \propto J_1J_6J_7$ and $J_3J_5J_6 \propto {\mathbb I}$.     
 The problem here is what to take for ``$i$,'' as the current $H\to J_1 m $ will take us out of the eigenspace of $N$.  But this is the problem we started with.   We {\it need \/} to double the space and keep  ``$i$''$=J_1$  and the real structure $\varphi =N$.   We are therefore retaining the ${\mathbb R}^{2r}$ Hilbert space. With $H=J_1m$ we have that  ${\mathcal T}=J_1J_6J_7$ commutes with$H$ and squares to $+{\mathbb I}$.

\item[D$\equiv \{{\rm O}(r)\times{\rm O}(r)\}/{\rm O}(r)\simeq {\rm O}(r)$:] Now the $m\in {\mathfrak m}_7$ anticommute with $J_8$ and, except for the factor ``$i$''$=J_1$,  we should stay in the space where $K=J_1J_2J_3$, $M=J_1J_4J_5$, $N=J_2J_4J_6$ and $P=J_1J_6J_7$ take the value $+1$
With $H=J_1 m$ we have that ${\mathcal C}=\varphi=N$ anticommutes with $H$, and brings us full circle.
\el

We have ended up with the symmetries displayed in table \ref{TAB:real_symmetries}.
\begin{table}
\begin{center}
\begin{tabular}{|c|c|c|c|c|c|}
\hline\hline
Cartan &TRS&PHS&SLS & Hamiltonian $M=G/H$ &   Classifying $Q$\\
\hline\hline
D&${\phantom-} 0$&$+1$&${\phantom-}0$& $ {\rm O}(16r)\times {\rm O}(16r)/{\rm O}(16r) \simeq {\rm O}(16r)$& $R_2$\\ 
DIII&$ -1$&$+1$& ${\phantom-}1$ &${\rm O}(16r)/ {\rm U}(8r)$ & $R_3$ \\
AII& $-1$& ${\phantom-}0$ &${\phantom-}0$& $  {\rm U}(8r)/{\rm Sp}(4r)   $ & $R_4$ \\
CII & $-1$&$-1$& ${\phantom-}1$&\{$ {\rm Sp}(4r)/{\rm Sp}(2r)\times {\rm Sp}(2r)\}   \times {\mathbb Z} $ & $R_5$   \\
C&$ {\phantom-}0$& $-1 $& ${\phantom-}0$& $ {\rm Sp}(2r)\times {\rm Sp}(2r) / {\rm Sp}(2r)\simeq {\rm Sp}(2r)   $& $R_6$ \\
CI& $+1$&$-1$& ${\phantom-}1$&$ {\rm Sp}(2r)/{\rm U}(2r)  $&$R_7$  \\
AI &$ +1$&${\phantom-}0$&${\phantom-}0$& $  {\rm U}(2r)/{\rm O}(2r) $& $R_0$   \\
BDI& $+1$&$+1$&${\phantom-}1$& $\{{\rm O}(2r)/{\rm O}(r)\times {\rm O}(r)\}  \times  {\mathbb Z} $& $R_1$  \\
D&$ {\phantom-}0$&$+1$&${\phantom-}0$& $ {\rm O}(r)\times {\rm O}(r)/{\rm O}(r)\simeq {\rm O}(r) $ &  $R_2$  \\
\hline\hline
\end{tabular}
\end{center}
\caption{The symmetries possessed by the Hamiltonians generating  each of the real Cartan spaces. The last column 
displays the space that classifies the bundle of fermionic ground states for that Hamiltonian class. It is one step further  in the Bott clock compared to that in table \ref{TAB:real_cartans}.}
\label{TAB:real_symmetries}
\end{table}

\section{Backwards and Forwards}
\label{SEC:involutions}

As mentioned in section \ref{SEC:symmetries},    the case of most interest, the Brillouin zone, requires us to take into account the ${\bf k}$ to $-{\bf k}$ effect of  antilinear symmetries.  In this short section we  make   a slight digression from the main story in order to introduce the  ingredients we will need to cope with this effect.

We  consider mutually  anticommuting real matrices  ${\tilde J}_i$ $i=1,\dots, k$ that now square to $+\mathbb I$. These obey
\be
{\tilde J}_i{\tilde J}_j+{\tilde J}_j{\tilde J}_i=2\delta_{ij}{\mathbb I},
\ee
and so form representations of the  Clifford algebra ${\rm Cl}_{0,k}$. 

 Again we seek the subgroups of $ {\rm O}(16r)$ that continue to commute with the ${\tilde J}_i$  as we enlarge the set.
An analysis similar to the one in the previous section gives  
\be
...{\rm O}(16r)\supset  {\rm O}(8r)\times {\rm O}(8r)\supset {\rm O}(8r)\supset {\rm U}(4r)\supset {\rm Sp}(2r) \supset  
{\rm Sp}(r)\times {\rm Sp}(r) 
 \supset {\rm Sp}(r)\supset {\rm U}(r) \supset {\rm O}(r)...
\ee 
The sequence of groups is ``backwards,''compared to the previous one,  but  the sequence of symmetric spaces
$$
R_0={\rm O}(16r)/{\rm O}(8r)\times {\rm O}(8r), \quad R_1= \{{\rm O}(8r)\times {\rm O}(8r)\}/{\rm O}(8r)\simeq O(8r), \quad R_2 ={\rm O}(8r)/{\rm U}(4r),
$$
 $$  R_3={\rm U}(4r)/{\rm Sp}(2r), 
\quad R_4= {\rm Sp}(2r)/ 
{\rm Sp}(r)\times {\rm Sp}(r), \quad R_5= 
\{{\rm Sp}(r)\times {\rm Sp}(r)\}/{\rm Sp}(r)\simeq  {\rm Sp}(r), 
$$
$$
R_6=\quad {\rm Sp}(r)/{\rm U}(r),\quad R_7={\rm U}(r)/{\rm O}(r)
$$
parametrizing the choice space for $\tilde J_{d+1}$ 
is in the same direction as before. It is, however, offset by two --- {\it i.e.\/}\ the set of  choices for ${\tilde J}_{d+1}$ is $R_{d}$. This  offset arises  because, once we have chosen ${\tilde J_1}$ and ${\tilde J_2}$ to bring us back to an orthogonal group, the choice of 
a higher $\tilde J$  is the same as choosing ${\tilde J_1} {\tilde J}_k$, and these square to $-{\mathbb I}$. The shift  is reflected in the Clifford algebra isomorphism ${\rm Cl}_{p,0} \otimes   {\rm Cl}_{0,2}\simeq {\rm Cl}_{0,p+2}$ given by
\bea
\tilde e_i &\leftrightarrow& e_i\otimes  \tilde e_1\tilde e_2, \quad i= 1,\dots,p, \quad \tilde e_1^2=\tilde e_2^2=1, \quad (\tilde e_1\tilde e_2)^2=-1,\nonumber\\
\tilde e_{p+1}&\leftrightarrow& {\rm id}\otimes  \tilde e_1,  \nonumber\\
\tilde e_{p+2}&\leftrightarrow&  {\rm id}\otimes e_2, \nonumber
\eea

Now suppose we have {\it both\/} $J_1$ and ${\tilde J}_1$. Then  ${\tilde J}_1$  decomposes  ${\mathbb R}^{16r}$  to  ${\mathbb R}^{8r}\times {\mathbb R}^{8r} $ and  ${\rm O}(16r)$ to ${\rm O}(8r)\times {\rm O}(8r)$, then $J_1$ breaks us down to the diagonal ${\rm O}(8r)$.   If we are given $J_1$ first, then ${\tilde J}_1$, we have ${\rm O}(16r)\to {\rm U}(8r)\to {\rm }O(8r)$, so we get to the same place independent of the order.    This observation reflects the Clifford algebra isomorphism ${\rm Cl}_{p,q} \otimes   {\rm Cl}_{1,1}\simeq {\rm Cl}_{p+1,q+1}$. (In \cite{kitaev_K}  Kitaev  says that the   positive Clifford generators effectively  ``cancel''  the negative ones.)   

 If we are given $q$ positive Clifford generators and $p$ negative ones,  and seek the degree of freedom to chose another {\it positive\/}  one, then that degree of freedom is given by $R_{q-p}$.  In other words, we end up and the group 
labelled by $q-p \,\hbox{\rm mod\,} 8 $ in the table below, and  degree of freedom is given by the coset of that group by the one to its right: 
 $$
\matrix{0&&1&&2&&3&&4&&5&&6&&7&&0
\cr{\rm O}&\to & {\rm O}\times {\rm O}& \to & {\rm O}& \to &{\rm U}& \to & {\rm Sp} & \to & {\rm Sp}\times {\rm Sp} &\to &{\rm Sp}&\to& {\rm U}&\to &{\rm O}\cr 
& R_0&& R_1&& R_2&& R_3&& R_4&& R_5&& R_6&& R_7&&}
$$
On the other hand, if we have $p$ negative generators and $q$ positive ones, the degree of freedom for the choice of the next  negative one is $R_{p-q+2}$.

%This is the origin of of Kitaev's
 %\be
 %\tilde {K}^{-q}_{\mathbb R}(\bar S^d)= \pi_0 (R_{q-d}).
 %\be 

\section{Real Representations}  
\label{SEC:real}

Much of the  insight into  topologically non-trivial band theory  has relied on illustrative  model Hamiltonians. These Hamiltonians are usually constructed from sets of gamma matrices, and the  time-reversal and charge-conjugation properties of the gamma matrices   play a key r{\^o}le.  
We therefore explore  the representation theory  of the irreducible $\gamma$-matrices that compose  the generally reducible $J_i$ and $\tilde J_i$ from the previous sections.   These matrices form representations of the Clifford algebras  ${\rm Cl}_{p,q}$ with $p$ generators $e_i$ obeying $e_i^2=-1$ and $q$ generators $\tilde e_i$ obeying $\tilde e_i^2=1$. The  ${\rm Cl}_{p,q}$ are {\it  real\/} algebras in that they consist of linear combinations of products of the generators with only real number co-efficients. (If we were to allow complex coefficients the distinction between $p$ and $q$ disappears.)  Although they are real algebras,  most physics discussions of their representation theory -- see for example \cite{polchinski} ---  consider representations over {\it complex\/} vector spaces.  This greatly obscures the connection between the algebra and the topology.  Atiyah Bott and Shapiro  showed \cite{atiyah_clifford} that the connection becomes much clearer when we  consider  irreducible representations (irreps) of   ${\rm Cl}_{p,q}$ over $\mathbb R$.    In other words,    $\gamma_i$'s that are $d_{p,q}$-by-$d_{p,q}$ real matrices.

The real representation theory of ${\rm Cl}_{p,q}$ is sharply different from the complex representation theory. In the complex case it is well known that  the irreps have dimension $2^{\lfloor( p+q)/2\rfloor}$. They are  unique when $p+q$ is even and there  are two inequivalent  irreps when $p+q$ is odd.  We might think that we can always get a real irrep from a complex one by simply  
using the correspondence (\ref{EQ:aplusib}) in reverse. If, however, the matrices all commute with a real structure then this process yields a reducible representation. Similarly some irreps  that are inequivalent  over the complex numbers become equivalent when expanded to become real. Consider the simple  example  ${\rm Cl}_{1,0}$. Over the complex numbers the two irreps $e_1\mapsto i$ and $e_1\mapsto -i$ are inequivalent, but when expanded to 
\be
e_1 \mapsto     \left(\matrix{0&-1\cr 1& \phantom - 0}\right), \quad \hbox{and}\quad e_1 \mapsto \left(\matrix{\phantom -0&1\cr -1&  0}\right)
\ee
we have 
\be
 \left(\matrix{\phantom -0&1\cr -1&  0}\right) = \left(\matrix{0&1\cr 1&0}\right)^{-1}\left(\matrix{0&-1\cr 1& \phantom - 0}\right) \left(\matrix{0&1\cr 1&0}\right),
 \ee
 so they are now  equivalent.

Consider first the Clifford algebra ${\rm Cl}_k\equiv {\rm Cl}_{k,0}$  generated by $k$ mutually anticommuting elements $e_i$ that square to $-1$. 
We find the results displayed in table \ref{TAB:real_irreps}.

The dimensions $d_k$ of the irreps are   found by characterizing the Clifford algebra as a matrix algebra over $\mathbb R$, $\mathbb C$ or $\mathbb H$ and multiplying the size of the matrices by one, two, or four, respectively.  This algebra is displayed in the second column, where, for example ${\mathbb H}(2)$ means the algebra of two-by-two matrices with quaternionic entries. After the fact,  we may realize that the dimensions could have  been read off  from table \ref{TAB:real_cartans} by noting that the product of $d_k$ with the  real, complex or quaternionic dimensions  of the commuting groups ---$8r$,  $4r$, $2r+2r$, $2r$, $2r$, $2r$, $1r+1r$, $r$ --- remains  constant at $16r$.

To see why there are sometimes two inequivalent  representations, consider the case    $p=3$,  $q=0$. The algebra is generated by $e_1$, $e_2$ and $e_3$ all of which square to $-1$. The product $\omega=e_1e_2 e_3$ commutes with everything in the algebra, and  $\omega^2=1$. In an irrep it can map to either $+\mathbb I$ or  $-\mathbb I$. No change of basis can change the sign, so there must be two inequivalent irreps.        

\begin{table}
 \begin{center}
\begin{tabular}{|c|c|c|c|c|}
\hline
$k$ & ${\rm Cl}_k$   &  $d_k$&  $N({\rm Cl}_k)$&  $N({\rm Cl}_k)/i^*N({\rm Cl}_{k+1})$  \\
\hline\hline
1 & ${\mathbb C}$ &2&  ${\mathbb Z}$ & ${\mathbb Z}_2$\\
2 & ${\mathbb H}$ & 4&${\mathbb Z}$ & $0$\\
3& ${\mathbb H}\oplus {\mathbb H}$ &4 & ${\mathbb Z}\oplus {\mathbb Z}$ & ${\mathbb Z}$\\
4& ${\mathbb H}(2)$ &8 & ${\mathbb Z}$ & $0$\\
5& ${\mathbb C}(4)$ &8& ${\mathbb Z}$ & $0$\\
6& ${\mathbb R}(8)$ &8& ${\mathbb Z}$ & $0$\\
7& ${\mathbb R}(8)\oplus {\mathbb R}(8)$ &8& ${\mathbb Z}\oplus {\mathbb Z}$ & ${\mathbb Z}$\\
8& ${\mathbb R}(16)$ &16 & ${\mathbb Z}$ & ${\mathbb Z}_2$\\
\hline\hline
\end{tabular}
\end{center}
\caption{The dimension and number of the real irreps of ${\rm Cl}_k$. Observe that $N({\rm Cl}_k)/i^*N({\rm Cl}_{k+1})$ coincides with $\pi_0(R_{k+1})$. }
\label{TAB:real_irreps}
\end{table}

%(The familiar corollary to Shur's lemma ---  that only scalar multiples of the identity commute with a set of irreducible matrices --- does not hold here, because $\mathbb R$ is not  an algebraically closed field.). 

The most interesting features in table \ref{TAB:real_irreps}  are the last two columns.   The symbol $N({\rm Cl}_k)$
denotes the  additive free group generated by the real irreps  of ${\rm Cl}_k$. If an algebra ${\rm A}$ has $d$ irreps $A_1, A_2,\ldots, A_d$ ,  then $N({\rm A})$   has elements 
\be
{\bf n}= n_1 A_1 \oplus  n_2 A_2\oplus  \ldots \oplus n_dA_d 
\ee
where the coefficients $n_i  \in {\mathbb Z}$, and  
\be
(n_1, n_2,\ldots, n_d)\in \underbrace{{\mathbb Z} \oplus {\mathbb Z}\oplus \ldots \oplus {\mathbb Z} }_{\hbox{$d$ copies}}.
\ee

When all  the coefficients are positive, we can think of these elements as being direct sums of irreps,  but it is convenient to allow the coefficients to be negative, so that     $N({\rm Cl}_k)$ becomes  a group.
Now the  natural inclusion $i:{\rm Cl}_k\to {\rm Cl}_{k+1}$ induces a  map \hbox{$i^*:N({\rm Cl}_{k+1})\to N({\rm Cl}_k)$} that consists of restricting each  irrep of ${\rm Cl}_{k+1}$ to ${\rm Cl}_k$. In  other words, we   simply omit  $\gamma_{k+1}$ and decompose the resulting (usually reducible) representation of ${\rm Cl}_k$ into its irreps.  Atiyah,  Bott and Shapiro  \cite{atiyah_clifford} then consider the quotient group $N({\rm Cl}_k)/i^*N({\rm Cl}_{k+1})$.  In this quotient group a (generally  reducible) representation maps to zero if it can be obtained  by restricting a representation of ${\rm Cl}_{k+1}$ to ${\rm Cl}_k$.

The origin of the ${\mathbb Z}_2$ groups in $N({\rm Cl}_k)/i^*N({\rm Cl}_{k+1})$ is easy to understand. Each time the dimension of the gamma matrices doubles, restriction to the smaller set of gamma matrices leads to a reducible representation that decomposes into two copies of the unique irrep of the smaller set. Thus an even number of copies of this irrep  maps to zero   in $N({\rm Cl}_k)/i^*N({\rm Cl}_{k+1})\simeq {\mathbb Z}_2$, while   an odd number of copies maps to to the non-trivial element $1\in{\mathbb Z}_2$. 

The $\mathbb Z$ groups arise in the cases when   ${\rm Cl}_k$ has two inequivalent irreps and  $N({\rm Cl}_k)={\mathbb Z}\oplus{ \mathbb Z}$. The restriction   of the unique irrep of  ${\rm Cl}_{k+1}$  then supplies a reducible representation of  ${\rm Cl}_k$ that contains one copy of each of its  inequivalent irreps.   In  $N({\rm Cl}_k)/i^*N({\rm Cl}_{k+1})$ we must therefore identify   $(n_1,n_2)\sim (n_1+m,n_2+m)$ for any integer $m$. The (Grothendiek) map
\be
(n_1,n_2) \to n_1-n_2,
\ee 
is then a group  isomorphism  taking each equivalence class to an integer.
Consequently  a representation that contains an equal number of the two 
inequivalent irreps maps to zero  in $N({\rm Cl}_k)/i^*N({\rm Cl}_{k+1})$, while one with an unequal numbers $n_1$, $n_2$ provides the element  $n_1-n_2\in {\mathbb Z}$.
 
 Observe from the table that, despite their very different origin,  $N(C_k)/i^*N(C_{k+1})$ coincides with $\pi_0(R_{k+1})$.  This is not by chance.  The real representation theory of the Clifford algebra is somehow capturing the topology of the classifying spaces. The way this works --- essentially  the Atiyah-Bott-Shapiro construction ---  will be the subject of the next section.

For ${\rm Cl}_{p,q}$ with  $p$ negative ($e^2=-1$)  and $q$ positive ($\tilde e^2=1$)   generators, the dimension $d_{p,q}$  of the irreducible real matrices is  displayed in table \ref{TAB:real_irreps2}. \begin{table} 
 \begin{center}
\begin{tabular}{|c||c|c|c|c|c|c|c|c|c|}
\hline
$d_{p,q}$ & q=0&1&2&3&4&5&6&7&8\\
\hline
p=0&1&1$_2$&2&4&8&8$_2$&16&16&16\\
1&2&2&2$_2$&4&8&16&16$_2$&32&\\
2&4&4&4&4$_2$&8&16&32&&\\
3&4$_2$&8&8&8&8$_2$&16&&&\\
4&8&8$_2$&16&16&16&&&&\\
5&8&16&16$_2$&32&&&&&\\
6& 8&16&32&&&&&&\\
7&8$_2$&16&&&&&&&\\
8& 16&&&&&&&&\\
\hline\hline
\end{tabular}
\end{center}
\caption{The dimension and number of the real irreducible representations  of ${\rm Cl}_{p.q}$.
The subscript $2$ denotes that there are two inequivalent   representations with this dimension.}
\label{TAB:real_irreps2}
\end{table}
 The entries in this  table can be extended down and to the right by noting that one step down and one step to the right doubles the size of the real matrices ---{\it i.e.\/}\  $d_{p+1,q+1}=2d_{p,q}$. This last fact follows from the isomorphism ${\rm Cl}_{p,q}\otimes {\rm Cl}_{1,1}\simeq{\rm Cl}_{p+1,q+1}$  given  by
\bea
e_i &\leftrightarrow& e_i\otimes  \tilde ee, \quad i= 1,\dots,p+q, \quad \tilde e^2=-e^2=1, \nonumber\\
\tilde e_{p+q+1}&\leftrightarrow& {\rm id}\otimes  \tilde e,  \nonumber\\
e_{p+q+2}&\leftrightarrow&  {\rm id}\otimes e, \nonumber
\eea
and the associated real-representation extension
\bea
\gamma_i &\to& \gamma_i\otimes \sigma_1, \quad i= 1,\dots,p+q,\nonumber\\
\tilde \gamma_{p+q+1}&\to&  {\mathbb I}\otimes \sigma_3,\nonumber\\
\gamma_{p+q+2}&\to&  {\mathbb I}\otimes (-i\sigma_2).\nonumber
\eea
We also have $d_{p+8,q}=d_{p,q+8}= 16d_{p,q}$.

Knowing the dimensions and number of the irreps  we can now read off  the groups $N({\rm Cl}_{p,q})/i^*N({\rm Cl}_{p+1,q})$. These are displayed in Table \ref{TAB:rep_groups}. Again observe that  $N({\rm Cl}_{p,q})/i^*N({\rm Cl}_{p+1,q})$ always coincides with $\pi_0(R_{p-q+1})$.

\begin{table}
\begin{center}
\begin{tabular}{|c||c|c|c|c|c|c|c|c|}
\hline
 & q=0&1&2&3&4&5&6&7\\
\hline
p=0&${\mathbb Z}_2$& ${\mathbb Z}$& $0$& $0$& $0$  & ${\mathbb Z}$& $0$ & ${\mathbb Z}_2$\cr
1&${\mathbb Z}_2$& ${\mathbb Z}_2$& ${\mathbb Z}$& $0$&$0$&$0$ & ${\mathbb Z}$ &$0$\cr
2& $0$&${\mathbb Z}_2$ &${\mathbb Z}_2$& ${\mathbb Z}$ &$0$ &$0$&$0$& ${\mathbb Z}$\cr
3& ${\mathbb Z}$&$0$& ${\mathbb Z}_2$&${\mathbb Z}_2$& ${\mathbb Z}$ & $0$&$0$&$0$\cr
4& $0$& ${\mathbb Z}$& $0$ & ${\mathbb Z}_2$&${\mathbb Z}_2$& ${\mathbb Z}$& $0$&$0$\cr
5&$0$&0&${\mathbb Z}$& $0$ &${\mathbb Z}_2$&${\mathbb Z}_2$&${\mathbb Z}$ &$0$\cr
6& $0$&0& $0$&${\mathbb Z}$ & $0$ &${\mathbb Z}_2$&${\mathbb Z}_2$&${\mathbb Z}$\cr
7&${\mathbb Z}$ &$0$& $0$& $0$& ${\mathbb Z}$ & $0$ &${\mathbb Z}_2$&${\mathbb Z}_2$\cr
\hline\hline
\end{tabular}
\end{center}
\caption{The quotient groups  $N({\rm Cl}_{p,q})/i^*N({\rm Cl}_{p+1,q})$. The table extends with period eight in both $p$ and $q$. Observe that $N({\rm Cl}_{p,q})/i^*N({\rm Cl}_{p+1,q})$ coincides with $\pi_0(R_{p-q+1})$.}
\label{TAB:rep_groups}
\end{table}

\section{\bf Constructing  representative  Hamiltonians} 
\label{SEC:abs_construction}

We now reach the pay-off  for  our labour in the previous sections. 
We use a  simplified  version of the Atiyah-Bott-Shapiro (ABS) construction \cite{atiyah_clifford,atiyah_KR} to generate model Hamiltonians in  any  given Altland-Zirnbauer symmetry class,  either over the $d$ sphere parametrized by a unit vector ${\bf x}$, or over  the   $d$-sphere equipped with  an  involution   that mimics the ${\bf k} \to -{\bf k}$ inversion  in the Brillouin zone. 
By applying   what we know about the real representations of  real Clifford algebras we show that the ABS  construction can provide Hamiltonians whose bundle of  negative energy  states lies in any of the  topological classes.

We desire to construct  operators  $\tilde Q({\bf x})$ and $\tilde Q({\bf k})$, that square to ${\mathbb I}$ (so that \hbox{$P({
\bf x})= (\mathbb I-\tilde Q({\bf x}))/2$} is the projection operator onto the negative eigenspace of $\tilde Q({\bf x})$) and  have the appropriate discrete symmetries  
\be
B_\eta \tilde Q^*({\bf x}) B_\eta^{-1} =\eta \tilde Q({\bf x}),
\ee
or 
\be
B_\eta \tilde Q^*({\bf k}) B_\eta^{-1} =\eta \tilde Q(-{\bf k}),
\ee
where $\eta=\pm 1$ and $B^*B= \pm {\mathbb I}$.  We  first construct Hamiltonians with  ${\bf x}\in S^d$. To do this we assemble  a set  of $d+1$ matrices  $\Gamma_n$ lying in the desired ${\mathfrak m}_{i}$ by setting  $\Gamma_n = J_{i+1}J_{i+1+n}$, $n=1,\ldots d+1$.  We know  from section \ref{SEC:symmetries} that when multiplied by a suitable operator ``$i$'' they possess  the required discrete symmetries.  The $\Gamma_n$  also obey the Clifford algebra
\be
\Gamma_n\Gamma_m+\Gamma_m\Gamma_n=-2\delta_{nm}\mathbb I,
\ee
so the normalized 
 linear combination 
\be
\tilde Q({\bf x}) = i\sum_{n=1}^{d+1} x_n \Gamma_n, \quad \|{\bf x}\|=1,
\label{EQ:abs}
\ee
squares to the identity and so   gives a map from $S^d$ to the classifying space $R_{i+3}$ (the space of choices for $J_{i+2}$ --- any  normalized linear combination of the $J_{i+1}^{-1}\Gamma_n$ provides  a candidate for $J_{i+2}$). 

Now any  bundle over a retractable space, such as a disc, is trivial. Conversely  a trivial bundle over a space $X$ can be extended to one over the retractable cone $CX$ over $X$ \footnote{The cone  $CX$ over $X$ is the retractable space $(X\times I)/( X\times\{0\})$ where $I=[0,1]$.}.
Consequently the  bundle of negative energy  states  will be  trivial or not depending on whether   $Q({\bf x})$  can be smoothly extended from $S^d$,  considered as the equator of $S^{d+1}$, to   the upper hemisphere of $S^{d+1}$ considered as the cone over $S_d$   ---taking care, of course,  to preserve  the discrete symmetries characterizing $\tilde Q$.
Thus,  whether  the bundle of negative energy states is trivial or not depends on whether the representation in which our $J_i$ (and hence the $\Gamma_n$)  live  extends to one with an additional  matrix $J_{i+3+d}$. In other words, whether  it  is  trivial or not in  $N({\rm Cl}_{i+d+2})/i^*N({\rm Cl}_{i+d+3})\simeq \pi_d(R_{i+3})\simeq \pi_0(R_{i+d+3})$.  That the equivalence classes of bundles are in one-to-one correspondence with  the elements of  $\pi_d(R_{i+3})\simeq \pi_0(R_{i+d+3})$ is exactly  the classification we obtained from the  purely topological considerations in   section \ref{SEC:bott_sequence}.  The construction in Eq.\ (\ref{EQ:abs})  is thus able to generate bundles in any equivalence class  and provides  a  one-to-one mapping between  the topological  $\tilde K(S^d)$ and  the purely algebraic  $N({\rm Cl}_{i+d+2})/i^*N({\rm Cl}_{i+d+3})$. This mapping  is  the  ABS  construction.

In the case of topological insulators, the base-space  $X$ is a Brillouin zone and therefore  a $d$-dimensional torus. We wish to obtain   ground state bundles over $X$ of the Bloch  Hamiltonians $H({\bf k})\equiv  \exp\{-i{\bf k}\cdot {\bf r}\}H  \exp\{i{\bf k}\cdot {\bf r}\}
$.  The antilinear maps $\mathcal T$ and $\mathcal C$  act as 
\be
{\mathcal T}H({\bf k}) {\mathcal T}^{-1} =H(-{\bf k}),\qquad {\mathcal C}H({\bf k} ){\mathcal C}^{-1} =-H(-{\bf k}).
\ee
Consequently, in addition to ordinary bundles over $X$, we need to study bundles that are compatible with the involution that takes ${\bf k}$ to $-{\bf k}$ (modulo reciprocal lattice vectors). This is the subject of {KR-theory\/} \cite{atiyah_KR}.  
We will not consider the complications due the fact that $X$ is torus, but instead 
model the Brillouin zone with its ${\bf k} \to -{\bf k}$ involution  by  replacing  the $d+1$ coordinates $x_n$  with  one co-ordinate $M$, and   $d$ coordinates $k_n$, $n=2,\ldots, d+1$. We can take these coordinates to obey 
\be
|{\bf k}|^2+M^2=1,
\ee   
so the ${\bf k}$-space has the topology a sphere with  the north pole $({\bf k}=0, M=1)$  representing the time-reversal-invariant $\Gamma$-point in the Brillouin zone, and with the south pole    \hbox{$({\bf k}=0, M=-1)$} being another fixed  point at infinity.

We also replace the $\Gamma_n$ that go with the $k_n$ by $\tilde \Gamma_n = J_{i+1}\tilde J_{i+1+n}$. These $\tilde \Gamma_n$ are real symmetric matrices obeying 
\bea
{\tilde \Gamma}_n{\tilde \Gamma}_m+{\tilde \Gamma}_m{\tilde \Gamma}_n &=&2\delta_{nm}\mathbb I, \quad 1< n,m\le d+1\nonumber\\
{\tilde \Gamma}_n{ \Gamma}_1+{\Gamma}_1{\tilde \Gamma}_n& =&0.
\eea
The $\tilde \Gamma_n$  have the same anticommutation relations with the  $J_i$ that are used to construct $\mathcal C$ and $\mathcal T$,  but because they do not get multiplied by ``$i$'' in constructing a Hamiltonian, they transform under these symmetries with an extra minus sign. Thus    
\be 
\tilde Q(M, {\bf k}) = i M \Gamma_1 + \sum_{n=2}^{d+1}k_n \tilde \Gamma_n, 
\ee
is a matrix that possesses the same $\mathcal C$ and $\mathcal T$ symmetries as $i \Gamma_1$. Because the $d$ positive generators 
take us backwards in the Bott clock, we  anticipate that the  topological classification  is  given by  $ \pi_0(R_{i+3-d})$ instead of $\pi_0(R_{i+3+d})$, but it is not exactly obvious how to see this from the homotopy perspective.  The algebraic  construction    makes it clear however:
The bundle  of negative energy states of $\tilde Q(M, {\bf k})$ will be trivial if and only if we can extend our Hamiltonian to one higher dimension. Because we do not wish to impose any relation between the upper and lower hemisphere of the $S^{d+1}$, the extra dimension should be of the non-inverting $x$ type. The extensibility of the bundle  is therefore determined by an element of   
\be
N({\rm Cl}_{i+2,d})/i^*N({\rm Cl}_{i+3,d}) =\pi_0(R_{i+3-d}).
\ee
This confirms our conjecture, agrees  with \cite{kitaev_K}, and reproduces table \ref{TAB:lud}.

\section{Conclusions}

We have explained the principle features of table \ref{TAB:lud} by relating them to the topology of the Bott sequence of Cartan symmetric spaces and to the real  representation theory of Clifford algebras.  Of course, most of the mathematics we have described is not new, being in \cite{bott,milnor,atiyah_clifford,atiyah_KR}. We hope, however, that  we have done a service   in explaining that the deep ideas in these papers have their origin in relatively simple facts from group theory and linear algebra. 

There are a number of issues that we have not discussed:  Firstly, we have given a simple algebraic  explanation  of the $d\to -d$  sign change effected by the ${\bf k}\to -{\bf k}$ inversion of the Block momentum. We have   not found a comparably simple homotopic  explanation of this  sign change. Such an explanation is desirable given its significance.    
Secondly, the spaces $\tilde K(X)$  have more than just the additive properties that arise from taking direct sums of fibres. They also have a ring structure that comes from taking tensor products.  The tensor product of two representations of a Clifford algebra is not a representation, however. To generate the ring structure one needs to consider graded
 representations and graded tenor products \cite{atiyah_clifford}. It remains to be seen if these concepts have a natural interpretation in the present context.  

It  would also be interesting to explore how the   groups   $N({\rm Cl}_{p,q})/i^*N({\rm Cl}_{p+1,q})$ are related to the pattern  of dimensional reduction in \cite{hughes1}.   These authors make use of complex gamma matrices, but whenever they commute with a suitable complex structure our real matrices can be rewritten as complex matrices of half the size. They cannot be so reduced   when they commute with a real structure, but in that case the complex matrices become real when written in a suitable basis.  Our arguments could, therefore, be recast into the language of complex gamma matrices, but at the cost of complicating the discussion.

We should add that the ABS construction has been used previously   in condensed matter physics  to characterize the stability of gapless Fermi surfaces \cite{horava}.  Our application is to gapped systems.

\section{Acknowledgements}

This work was supported by the National Science Foundation  under grant NSF DMR 09-03291.  MS would like to thank Matthew Ando and Sheldon Katz for   explaining K theory, and Taylor Hughes for not only patiently explaining  the physics of topological insulators, but also for help with the present exposition.

\end{document}